# Tunable Structural Transmissive Color in Fano-Resonant Optical Coatings Employing Phase-Change Materials

Yi-Siou Huang,[a,b] Chih-Yu Lee,[a] Medha Rath,[c] Victoria Ferrari,[a,b] Heshan Yu,[a,d] Taylor J. Woehl,[e] Jimmy Ni,[f] Ichiro Takeuchi,[a,g] Carlos Ríos,[a,b,*]

[a] Department of Materials Science and Engineering, University of Maryland, College Park, MD 20742, USA
[b] Institute for Research in Electronics and Applied Physics, University of Maryland, College Park, MD 20742, USA
[c] Department of Chemistry and Biochemistry, University of Maryland, College Park, MD 20742, USA
[d] School of Microelectronics, Tianjin University, Tianjin 300072, China
[e] Department of Chemical and Biomolecular Engineering, University of Maryland, College Park, MD 20742, USA
[f] U.S. Army, Combat Capabilities Development Command, Army Research Laboratory
[g] Maryland Quantum Materials Center, Department of Physics, University of Maryland, College Park, MD 20742, USA
* Corresponding author: riosc@umd.edu

**Abstract:**
Reversible, nonvolatile, and pronounced refractive index modulation is an unprecedented combination of properties enabled by chalcogenide phase-change materials (PCMs). This combination of properties makes PCMs a fast-growing platform for active, low-energy nanophotonics, including tunability to otherwise passive thin-film optical coatings. Here, we integrate the PCM $Sb_2Se_3$ into a novel four-layer thin-film optical coating that exploits photonic Fano resonances to achieve tunable structural colors in both reflection and transmission. We show, contrary to traditional coatings, that Fano-resonant optical coatings (FROCs) allow for achieving transmissive and reflective structures with narrowband peaks at the same resonant wavelength. Moreover, we demonstrate asymmetric optical response in reflection, where Fano resonance and narrow-band filtering are observed depending upon the light incidence side. Finally, we use a multi-objective inverse design via machine learning (ML) to provide a wide range of solution sets with optimized structures while providing information on the performance limitations of the PCM-based FROCs. Adding tunability to the newly introduced Fano-resonant optical coatings opens various applications in spectral and beam splitting, and simultaneous reflective and transmissive displays, diffractive objects, and holograms.

## 1. Introduction

Thin-film optical coatings are crucial in manipulating the reflection, absorption, and transmission spectra in all sorts of objects onto which they are conformally deposited [1,2]. Optical coatings have been employed commercially in a plethora of applications, namely anti-reflective coatings [3,4], color pixels [5], paintings [6], polarizing filters [7], narrow and broadband filters [8,9], beam splitting [10], dielectric mirrors [11], density filters [12], etc. All sorts of optical phenomena in slabs, periodic, or cavity-like resonant structures have long been studied for coatings, levering the optical properties of different materials and metamaterials [13]. However, only recently, a new optical coating was proposed by ElKabbash *et al.*[14], which employs Fano resonances in low-dimensional thin-film stacks comprising two coupled cavities: a transparent low-index material sandwiched between two ultrathin metal films and an absorptive, large refractive index thin film. This structure is equivalent to combining strong interferences effects in a broadband absorber [15] and a narrow-band filter that uses a dielectric Fabry-Perot cavity [2,16]. Provided the precise combination of geometrical and material properties, the resonant wavelength of both cavities can couple and resonate in what is known as the Fano resonance [17]. The most interesting feature of this Fano-resonant optical coating (FROC) is enabling reflectance and transmittance peaks at the same wavelength while achieving vivid structural colors despite using a broadband absorber.

FROCs, like most coatings, have been demonstrated using passive materials for a single optical response, while few other active coatings for structural color have relied on volatile phenomena, such as electrochromism [18,19], microelectromechanical modulation [20], liquid crystals [21], volatile phase change materials [22,23], among others. Applications with fast response but slow or sporadic tunability undergo suboptimal energy performance due to the volatile nature of such platforms, which require a constant power supply to hold a desired color. Phase-change materials (PCMs) are materials capable of filling this gap by introducing a zero-static power approach for nonvolatile optical modulation [24]. This is achieved via a solid-to-solid phase transformation between the amorphous and the crystalline states of chalcogenide materials, which are both stable and optically distinct with unprecedented refractive index contrasts ($\Delta n < 3.5$, $\Delta k < 2.5$). PCMs have already been used to demonstrate various structures for color filtering in reflection [5,16,25] and transmission [26], and metasurfaces [27–29].

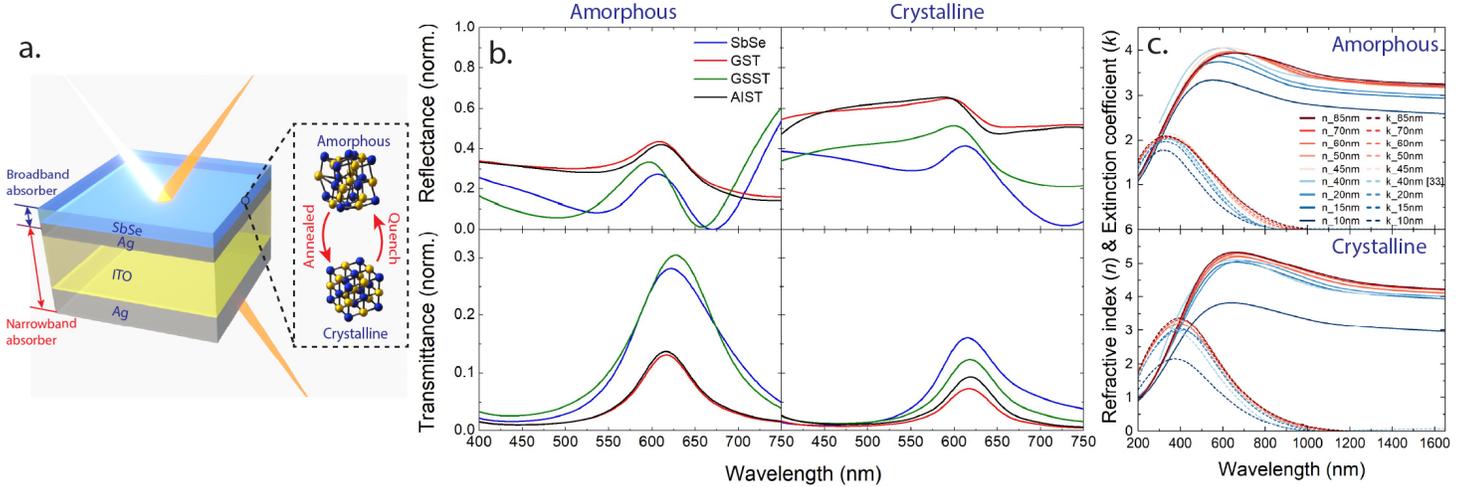

**Figure 1.** PCM-FROC device (a) Schematic of the PCM – FROC structure (b) Reflectance and transimisttance of 17.3nm amorphous $Sb_2Se_3$ (13.3nm crystalline $Sb_2Se_3$)/ 30nm Ag/ 105nm ITO / 25nm Ag on a fused silica substrate in the visible spectrum. (c) $Sb_2Se_3$ refractive index and extinction coefficient dependence on film thickness.

Here, we demonstrate PCM-based Fano resonant optical coatings (PCM-FROCs) for tunable structural colors in uniform thin-film coatings and micropatterned structures. We demonstrate that judicious thin-film engineering determines the reflected and transmitted color within the visible spectrum, while the phase transition of $Sb_2Se_3$ allows for amplitude and resonance wavelength modulation. Furthermore, we use machine learning (ML) techniques to accelerate the inverse design of PCM-FROCs, in a similar fashion that other free-space nanophotonic devices [30–32]. In particular, we use a genetic algorithm to optimize PCM-FROCs' transmission with purer color at a wavelength of interest, focusing on peaks with high transmittance and narrower full width at half maximum (FWHM). With this approach, a series of solution sets is generated efficiently from a large search space; an otherwise time-consuming task if choosing manually a handful of free parameters based on analytical theories to perform simulations and experiments. Moreover, we use ML to study the spectral limitations of PCM-FROCs.

## 2. Material and methods

*2.1 PCM Fano-resonant optical coating*

The PCM-FROC is a simple four-layer thin-film stack, as shown in **Fig. 1a,** that couples two optical resonators[14]. The first resonator on the top is a strongly damped system consisting of a broadband absorber onto a metallic layer, $Sb_2Se_3$ and Ag, respectively. The second resonator is a Fabry-Perot cavity with a narrowband filtering response using a typical metal/lossless material/ metal stack, which corresponds to Ag/ITO/Ag stacks in our device. We choose ITO because it is a transparent conductive material suitable for our structural color approach while enabling electrical methods to switch the PCMs actively [5]. By stacking one structure on top of the other, a Fano resonance raising from the coupling between both cavities leads to an unprecedented peak, at the same wavelength, in reflection and transmission.

*2.2 Phase-change materials (PCMs)*

To confine light within the PCM while achieving broadband absorption, we require both a material with a large refractive index ($n$) and a significant extinction coefficient ($k$). To find an optimum alloy for transmission, which is the main focus of this work, we compared via computational simulations the performance of four of the most typical PCMs with absorption in the entire visible spectrum, namely $Sb_2Se_3$ (SbSe), $Ge_2Sb_2Te_5$ (GST), $Ge_2Sb_2Se_4Te$ (GSST), and $Ag_3In_4Sb_{76}Te_{17}$ (AIST). We used Finite-Difference Time-Domain (FDTD) simulations on ANSYS Lumerical™ and the refractive indices plotted in the supplementary **Fig. S1**. **Fig. 1b** shows the simulation results for 17.3nm PCM/ 30nm Ag/ 75nm ITO/ 25nm Ag stacks on a fused silica substrate. Comparing the transmittance peaks, $Sb_2Se_3$ shows the best performance among the four PCMs, which we attribute to its large $n$, especially in the crystalline state, combined with a lower $k$. This combination of properties leads to a Fano resonance with lower losses in transmission, but also allows a clear peak in reflection, as opposed to a predominant broadband absorber response—the case for GST and AIST. Based on these results, we chose $Sb_2Se_3$ as our PCM platform and, consequently, characterized its optical properties as a function of thickness, given the ultra-thin films used in our PCM-FROC structures. **Fig. 1c** shows the refractive index and extinction coefficient for amorphous and crystalline $Sb_2Se_3$ films using ellipsometry (see Section 2.3), which agrees well with previous literature [33]. Both $n$ and $k$ are thickness dependent, decreasing for thinner layers. This effect is considered in the spectral simulations of PCM-FROCs, especially given that $Sb_2Se_3$ films shrink upon crystallization.

*2.3 Sample fabrication and characterization*

To characterize the optical properties of $Sb_2Se_3$, we sputtered films with different thicknesses onto Si substrates using an AJA Orion-3 UHV Sputtering System at room temperature. We measured the complex refractive index using a J.A. Woollam M-2000D Spectroscopic Ellipsometer and performed a Tauc–Lorentz fitting from the general oscillator (Gen-Osc) model. To attain the crystalline state of the PCM-FROC device, we annealed the samples on a hotplate at 200°C for 10 minutes in a nitrogen environment to prevent oxidation. To fabricate the PCM-FROC multi-layer thin-film structures, we deposited Ag and ITO layers using an AJA ATC 1800 Sputtering System at room temperature. Before depositing the top $Sb_2Se_3$ layer, the reflectance of the samples was measured. ITO thicknesses were calibrated by fitting the Ag/ITO/Ag measured spectra to calculations using a transfer matrix algorithm [34]. Lastly, $Sb_2Se_3$ was deposited following the same process described above. The reflectance spectra were measured with using a home-built system featuring a HORIBA JOBIN YVON VS140 linear array spectrometer. For transmission measurements, the samples were attached to a modified Thorlabs FOFMS cuvette holder for fiber-coupled transmittance measurement. To fabricate the samples with patterned logos, we used a Heidelberg MLA150 Maskless aligner to transfer the patterns onto the positive photoresist, followed by the deposition via sputtering of Ag/ITO/Ag with different ITO thicknesses by using a combination of shadow masks. We then performed a lift-off process to release the photoresist and deposited $Sb_2Se_3$ onto the entire surface.

*2.4 Machine learning*

The multi-objective optimization was constructed with the Pymoo package [35] in combination with optical spectra generated by our Python-scripted transfer matrix algorithm [34]. We varied the thickness of the four layers from 1 nm to 150 nm: $Sb_2Se_3$, Ag, ITO, and Ag respectively. To obtain results of Fano-resonance and peaks within the preferred range (350-750 nm), several constraints were implemented. An important constraint was that transmissive devices, the only ones studied with the ML algorithm, should display both reflection and transmission peaks at the same or very close wavelengths; this way, we guaranteed only Fano resonances and avoided Fabry-Perot solutions from Ag/ITO/Ag structures with vanishing PCM thicknesses. The number of objectives varies from case to case; some include the maximization of the peak transmittance, the maximization of peak shift upon $Sb_2Se_3$ phase switching, and the minimization of the FWHM. The parameters used in the non-dominated sorting genetic algorithm are as follows: The population size was set at 500. Only integers are randomly sampled. The algorithm generates 500 offspring combinations for each generation. The probability of crossover and mutation are both 80%. Elimination of duplicated offspring is included. Violation of constraints is added to be a penalty of the optimization process. Data and algorithms are available on Github [34].

## 3. Results and discussion

*3.1 Reflective-only PCM-FROC*

We first demonstrate a reflective-only PCM-FROC multi-layer thin film structure consisting of 9.5nm $Sb_2Se_3$/ 12nm Ag/ $t$ ITO/ 100nm Ag (from the top to bottom layer) on Si substrate, as shown in **Fig. 2a**. The bottom Ag layer works as a mirror in the reflective structure; typically, a thickness above 50 nm will achieve this function. Furthermore, by varying the thickness $t$ of the ITO layer, different reflectance spectra and, thus, colors are obtained, like previously demonstrated PCM-based Fabry-Perot structures [5,16]. In **Fig. 2b**, we present the simulated color swatch for both the amorphous and the crystalline states of the PCM-FROC devices using the XYZ tristimulus calculated from the simulated reflectance spectrum [36]. **Fig. 2c** shows a picture of the fabricated devices, displaying excellent agreement with the simulated colors in **Fig. 2b.** Moreover, we demonstrate a good agreement between the measured spectra and the FDTD simulations for three different ITO thicknesses, shown in **Fig. 2d**. Since the $Sb_2Se_3$ film undergoes a volumetric contraction upon crystallization, both layer thicknesses were calibrated before and after annealing using ellipsometry (see *Section 2.3*). We found that the $Sb_2Se_3$ layer shrank from 9.5 nm to 8.0 nm after annealing, and, generally across the multiple thicknesses explored in this work, between 15% and 20%. Further engineering can lead to narrower reflection peaks, for instance, by using thicker absorptive media. [14]

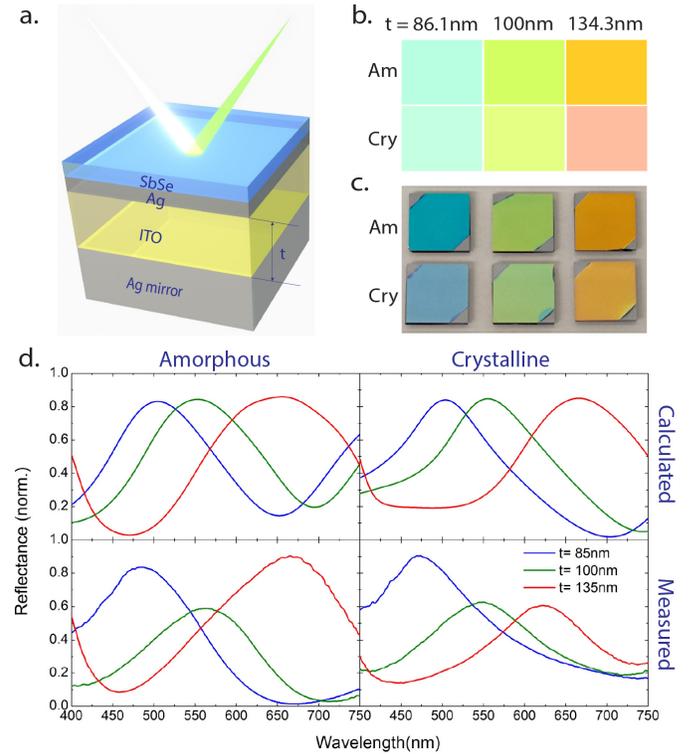

**Figure 2. Reflective-only PCM-FROC device. (a)** Schematic of the multi-layer reflective-only PCM-FROC structure consisting of 9.5nm $Sb_2Se_3$/ 12nm Ag/ t ITO/ 100nm Ag. **(b)** Simulated color swatches for different ITO and in both amorphous (Am) and crystalline (Cry) states. **(c)** Pictures of reflective-only PCM-FROCs with different ITO thicknesses and in both phase states. **(d)** Comparison of the measured and the FDTD simulated spectra in both the amorphous and crystalline states for the reflective-only PCM-FROC.

While the PCM-FROCs in the amorphous state match well the simulation results, there is a discrepancy between the calculated and the measured spectra for the crystalline samples. We attribute this effect to the surface variations in the Ag thin films, which, at the same annealing conditions we used to crystallize $Sb_2Se_3$, undergo dewetting, thus affecting its optical properties [37]. To understand the effect of the annealing process in our Fabry-Perot cavity, we measured the reflection spectra of similar Ag/ITO/Ag stacks before and after 10 min on a hotplate at 200°C. In supplementary **Fig. S2**, we show how the spectral response for three different stacks changes after the annealing process, affecting, in particular, the response at longer visible wavelengths. This undesired effect is more pronounced in the transmissive samples studied in *Section 3.2* and can be suppressed by employing metals with better thermal stability or by doping the silver film [37]. Temperature-independent optical response for all the materials involved (except, clearly, the PCM) is imperative in future efforts to reversibly switch PCM-FROCs, since re-amorphization stimulus requires melting temperatures above 600°C.

*3.2 Transmissive PCM-FROC*

We now modify the multi-layer thin film structure to demonstrate transmissive PCM-FROC devices consisting of 17.3nm $Sb_2Se_3$/ 30nm Ag/ *t* ITO/ 25nm Ag, although, in supplementary **Fig. S3**, we show a larger parametric exploration varying both the ITO and the $Sb_2Se_3$ thicknesses. To achieve a transmissive structural color, the thickness of the bottom silver layer is reduced, and silicon is replaced by a transparent fused silica substrate. **Fig. 3a** and **Fig. 3b** show the spectral comparison and pictures of transmissive PCM-FROC devices with different ITO thicknesses in both the amorphous and the crystalline states, displaying a good agreement between measured and calculated spectra. In these samples, the $Sb_2Se_3$ film thickness reduces on average from 17.3 nm to 13.3 nm after annealing, which is accounted for in the spectra simulations.

Besides structural color in transmission, PCM-FROC devices display drastically distinct reflection depending on the side of the sample onto which the light is incident. Such asymmetry results from the location of both resonators, and the interaction of the light with each of them. In **Fig. 3c**, we compare the calculated and measured reflectance spectra when the light incidence takes place on the $Sb_2Se_3$, displaying good agreements for each of the different thicknesses of ITO. By comparing **Fig. 3a** and **Fig. 3c**, we note that both the reflective and the transmittance peaks overlap, i.e. incidence on the PCM side displays the expected Fano resonance [14]. This response is unlike most color filters, where reflectance peaks usually accompany transmittance deeps. The same PCM-FROC structures show different color in reflection when light is incident from the rear side, i.e. from the silica substrate. In this case, the stack responds as a narrow-band filter meaning that the light predominantly resonates within the ITO cavity (see Supplementary **Fig. S4**), leading to Fabry-Perot type spectra, as shown in **Fig. 3e** and **Fig. 3f**. The transmission spectrum, as expected, does not undergo this asymmetric effect since light interacts with both resonators regardless the side of incidence.

Furthermore, we demonstrate in **Fig. 4a** the contrast between a PCM broadband absorber and micropatterned PCM-FROCs on the same fused silica substrate. The Army Research Laboratory and the University of Maryland (UMD) logos were patterned onto quartz samples following the fabrication process described in *Section 2.4*. The result is logos featuring PCM-FROCs with a background corresponding to amorphous $Sb_2Se_3$ on fused silica acting as a broadband absorber. The different structural colors for the logos in **Fig. 4a** and **Fig. 4b** correspond to different thicknesses of ITO (70 nm, 90 nm, and 110 nm). A zoom-in image of one of the patterns under an optical microscope with transmission illumination is shown in **Fig. 4c**. As expected, the transmittance of the UMD logo in **Fig. 4b** (top image) shows the same colors as the reflectance of the UMD logo (**Fig. 4b** middle image). Moreover, **Fig. 4b** (bottom) shows the narrow-band filter reflection upon rear-side illumination, which leads to a completely different perceived color. The supplementary video shows the different colors of the UMD logos sample while rotating, showing the color robustness upon large angle incidence, which was previously demonstrated in [14].

*3.3 Machine Learning inverse design*

In transmissive PCM-FROCs, the ideal structural color response should feature high transmittance and narrow FWHM for purer colors. Upon switching the phase-change material, in addition, the ideal situation is to achieve either a pronounced difference in the transmittance peak or the peak's wavelength, depending on the targeted application. Optimizing one or several of these desired variables simultaneously is a difficult task if done manually due to the considerable number of variables that create a large search space. Instead, we used the ML algorithm described in *Section 2.4* to find the optimum solutions depending on different objectives. The first case we demonstrate in **Fig. 5a** is the simultaneous maximization of the transmittance peak and minimization of the FWHM, for any wavelength, in PCM-FROCS with amorphous Sb$_2$Se$_3$. After finding the different structures, a simulation was performed to calculate the response of the same structure in the crystalline state, which is also plotted in **Fig. 5a**. In this case study, we observe that the two objectives and the wavelength are all directly proportional, which creates an undesired effect and shows the limitations of PCM-FROCs since a large peak transmittance with a minimal FWHM is unattainable. The spectrum with the best combination of results (i.e., closer to the optimal solution: minimum FWHM and maximum transmittance) shows a transmission peak at 430 nm, with a normalized intensity of approximately 0.23 (23% transmission) and FWHM of 60 nm. The spectrum with the highest transmittance displays a transmission peak at 720 nm with 40% of transmitted light but at the cost of broadening the FWHM to 120 nm. We also Achieving a large shift in the transmission peak upon phase switching would be desirable in spectrum-splitting applications. However, in supplementary section S7, a three objective optimization including the transmission peak amplitude, the FWHM, and the peak shift reveals that PCM-FROCs display small spectral shifts upon $Sb_2Se_3$ phase transition, where ~20 nm shifting is the maximum observed (see supplementary **Fig. S7**). Given this constraint, our current four-layer stack PCM-FROC is more suitable for amplitude-switching applications, like tunable beam splitting. Accordingly, the following optimization note that as the peak locates at a longer wavelength, the contrast of the peak transmittance in the two states is larger, because of this effect, the optimal solutions suggest thicker PCM layers. In supplementary **Fig. S5**, we show the histogram of thickness distribution from the population generated by ML for **Fig. 5a**. The two layers of Ag are both around 20 nm for optimal results, the PCM thickness varies but near 30nm suggest the largest transmittance contrast, and ITO tends to be thicker over 120 nm to allow high transmission. As ITO grows thicker, a higher-order Fano-resonance peak appears at a short wavelength, which is characterized by smaller transmittance and FWHM, as shown in **Fig. 5a**. While the goal is to optimize the spectra of the PCM-FROC in both $Sb_2Se_3$ states, we found that the optical responses in amorphous and crystalline states have a positive correlation. Therefore, optimizing in one phase provides accurate results to investigate the overall transmission properties in both of them. This is evident by comparing the similar results achieved in **Fig. 5a** with two objectives (FWHM and peak transmitance for amorphous $Sb_2Se_3$), and supplementary **Fig. S6**, where four objective optimization was performed (FWHM and peak transmittance for amorphous and crystalline $Sb_2Se_3$.)

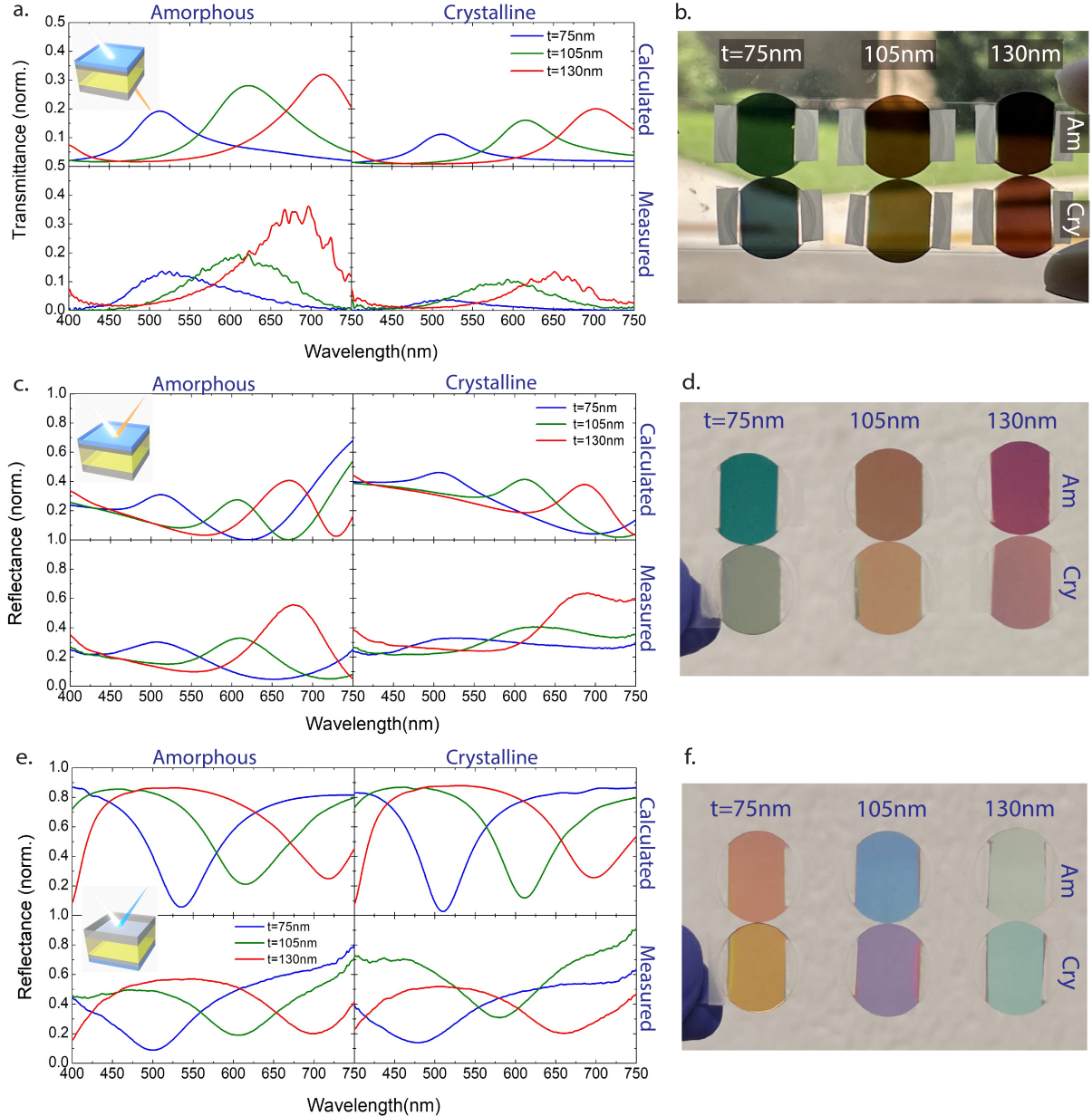

**Figure 3. Transmissive PCM-FROCs with varying ITO thicknesses: 17.3nm Sb$_2$Se$_3$/ 30nm Ag/ *t* ITO/ 25nm Ag for t= 75 nm, 105 nm, and 130 nm. (a)** Comparison between the measured and the FDTD simulated transmittance spectra in both the amorphous and crystalline states. **(b)** Picture of the transmissive PCM-FROC device with different ITO thicknesses. **(c)** Comparison between the measured and the FDTD simulated reflectance spectra in the amorphous and crystalline states under illumination on the Sb$_2$Se$_3$ side. **(d)** Picture of the PCM-FROC samples in reflection with illumination on the Sb$_2$Se$_3$ side. **(e)** Comparison of the measured and the FDTD simulated reflectance spectra in the amorphous and crystalline states under rear-side illumination on the substrate side. **(f)** Picture of the PCM-FROC samples under illumination from the substrate side.

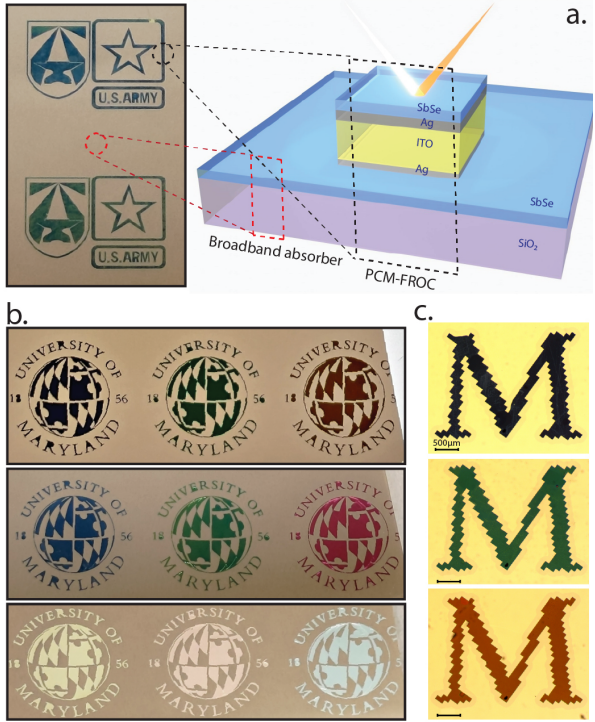

**Figure 4. Micropatterned PCM-FROCs. (a)** Schematic and image for both the broadband absorber and the PCM-FROC on the same fused silica substrate **(b)** Images for transmissive UMD logos with different ITO thicknesses. Top: color in transmission. Middle: color in reflection (from $Sb_2Se_3$ side). Bottom: color in reflection from the substrate side (horizontally mirrored for comparison). **(c)** Zoom in to the transmissive blue, green, and red images in (b). The scale bar corresponds to 500 μm.

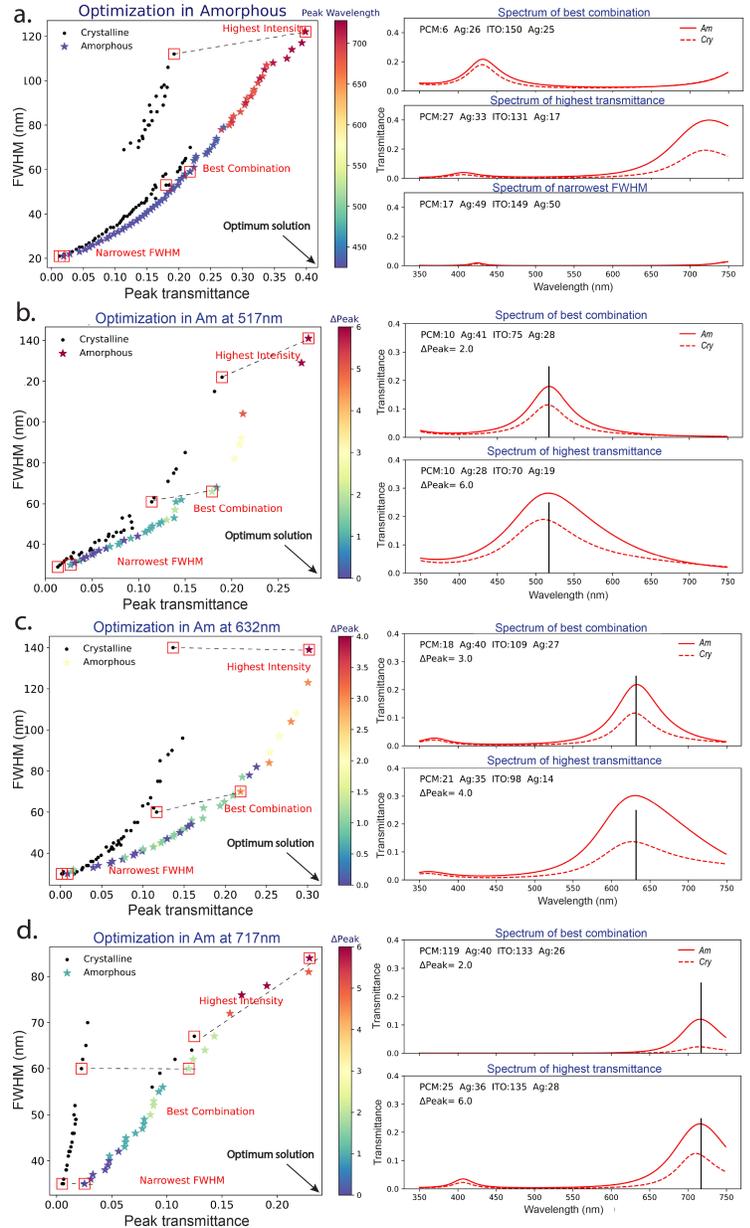

**Figure 5. ML-optimized PCM-FROCs (a)** Two-objective optimization for transmittance peaks and FWHM in the amorphous state only. The crystalline data are calculated for the structure optimized performed in the amorphous state. **(b)-(d)** Optimization for maximum transmittance and minimum FWHM in the amorphous state for three wavelengths: 517nm, 632nm, and 717nm, same as in Fig. 3. The spectra on the right panels correspond to the solution sets highlighted on the left panels.

Table 1. Comparison between experimental results and ML-optimized results

| | | | | Amorphous | | Crystalline | |
|---|---|---|---|---|---|---|---|
| | Wavelength in Amorphous | Thickness (nm) | Peak shift (nm) | Peak Transmittance | FWHM (nm) | Peak Transmittance | FWHM (nm) |
| **Exp** | 517 nm | 17.3/ 30/ 75/ 25 | 1 | 0.18 | 91 | 0.10 | 76 |
| | 632 nm | 17.3/ 30/ 105/ 25 | 7 | 0.25 | 114 | 0.15 | 88 |
| | 717 nm | 17.3/ 30/ 130/ 25 | 16 | 0.27 | 100 | 0.17 | 109 |
| **ML** | 517 nm | 10/ 41/ 75/ 28 | 2 | 0.18 | 66 | 0.11 | 61 |
| | 632 nm | 18/ 40/ 109/ 27 | 3 | 0.22 | 70 | 0.12 | 60 |
| | 717 nm | 25/ 36/ 135/ 28 | 1 | 0.23 | 81 | 0.11 | 61 |

cases do not include this objective to save computational costs.

An advantage of ML is its capability to precisely find solutions with target wavelength for the transmitted color. Three different wavelengths, the same as achieved experimentally in **Fig. 3**, are benchmarked in **Fig. 5b-d**: 517 nm, 632 nm, and 717 nm, optimized in the amorphous states with only two objectives. The results agree with the parameters used in the experiments, particularly the thickness of ITO, which is considered the most significant layer in the color response. **Table 1** also lists the comparison between experimental and ML-optimized results. ML combined with transfer matrix modeling is advantageous in complicated multi-objective problems through quantitative analysis and allows for flexibility in modifying algorithms to meet practical requirements, for instance, the ~20% shrinkage of PCM upon crystallization and thickness dependence of the PCM optical properties.

## 4. Conclusion

Chalcogenide phase-change materials are a versatile platform that enables nonvolatile tuning in various nanophotonic structures. In this work, we have demonstrated their integration into novel Fano-resonant optical coatings to allow for tunable structural color in transmission and reflection. Moreover, we have used a Machine Learning approach to find optimal structures and study the limitations of PCM-FROCs with different multi-objective optimization searches. In particular, we have demonstrated that $Sb_2Se_3$ is the optimum alloy to build PCM-FROCs in the visible spectrum, given its large refractive index in the crystalline state, the refractive index contrast between both states, and its moderate extinction coefficient. Moreover, we have optimized transmissive tunable color structures, a difficult task using PCMs due to their high optical losses. In comparison with the only other demonstration of PCM in transmission color [26], which uses a more complex six-layer structure, PCM-FROCs display transmission peaks with smaller FWHM and zero transmission at wavelengths far from the Fano resonance, i.e., the transmitted color is purer and without the pale perception due to other undesired wavelength contributions. However, PCM-FROCs suffer from low transmittance, up to 40%, due to the nature of the Fano resonance, which enables both reflection and transmission at the same resonant wavelength, thus, splitting the incoming light intensity. Additionally, PCM-FROCs in the current four-layer configuration displays a small peak shift which needs further optimization for applications in spectral splitting.

Furthermore, we demonstrated an unique asymmetric response of our PCM-FROC structure, which allows for three colors from each structure. The front-side (PCM side) reflection and the transmission display Fano-resonant colors, while the rear-side (fused silica side) reflection displays a narrow-band behavior since light resonates stronger with the Fabry-Perot ITO cavity than with the PCM thin film. The multiplicity of colors in these structures can be used in various applications, including trichroic optical filters, spectral and beam splitting, encryption, multiplexed holography, and others. Finally, our approach can be seamlessly integrated with demonstrated active electro-thermal switching approaches, especially given that the PCM sits directly on top of a metal layer that can be patterned into a microheater. [38–41]

## Acknowledgements

C.R. acknowledges support from the U.S. National Science Foundation under Grant ECCS-2210168 and the Minta Martin Foundation through the University of Maryland. I.T. is supported by ONR MURI N00014-17-1-2661. M.R. and T.J.W. acknowledge support for this work from the National Science Foundation under grant NSF-CBET-2025249.

## Data availability

The raw data required to reproduce these findings are available upon reasonable request from the authors. The ML algorithm and the raw data from the optimization runs are available to download from [34].

# Supplementary material

# Structural Color in Tunable Transmissive Fano-Resonance Optical Coatings Employing Phase Change Materials

Yi-Siou Huang,[a,b] Chih-Yu Lee,[a] Medha Rath,[c] Victoria Ferrari,[a,b] Heshan Yu,[a,d] Taylor J. Woehl,[e] Jimmy Ni,[f] Ichiro Takeuchi,[a,g] Carlos Ríos,[a,b,*]

[a] Department of Materials Science and Engineering, University of Maryland, College Park, MD 20742, USA
[b] Institute for Research in Electronics and Applied Physics, University of Maryland, College Park, MD 20742, USA
[c] Department of Chemistry and Biochemistry, University of Maryland, College Park, MD 20742, USA
[d] School of Microelectronics, Tianjin University, Tianjin 300072, China
[e] Department of Chemical and Biomolecular Engineering, University of Maryland, College Park, MD 20742, USA
[f] U.S. Army, Combat Capabilities Development Command, Army Research Laboratory
[g] Maryland Quantum Materials Center, Department of Physics, University of Maryland, College Park, MD 20742, USA
* Corresponding author: riosc@umd.edu

This supplementary information contains the following sections:

**S1.** Comparison of four PCMs refractive index
**S2.** Effect of annealing process on the optical response of Ag/ITO/Ag
**S3.** RGB color simulation of PCM-FROC devices
**S4.** FDTD simulations of the electric field in the PCM-FROC
**S5.** Histogram of thickness distribution
**S6.** Four objectives optimization: FWHM and peak transmittance in Amorphous and Crystalline states
**S7.** Three objectives optimization: FWHM, peak transmittance and peak shift
**References**

## S1. Comparison of four PCMs refractive index

**Fig. S1** shows the refractive index of the chalcogenide phase-change materials $Sb_2Se_3$ (SbSe), $Ge_2Sb_2Te_5$ (GST), $Ge_2Sb_2Se_4Te$ (GSST), $Ag_3In_4Sb_{76}Te_{17}$ (AIST) used in the simulations of Fig. 1 in the main text. SbSe and GSST were measure using ellipsometry. GST and AIST are re-plotted from reference [1], respectively.

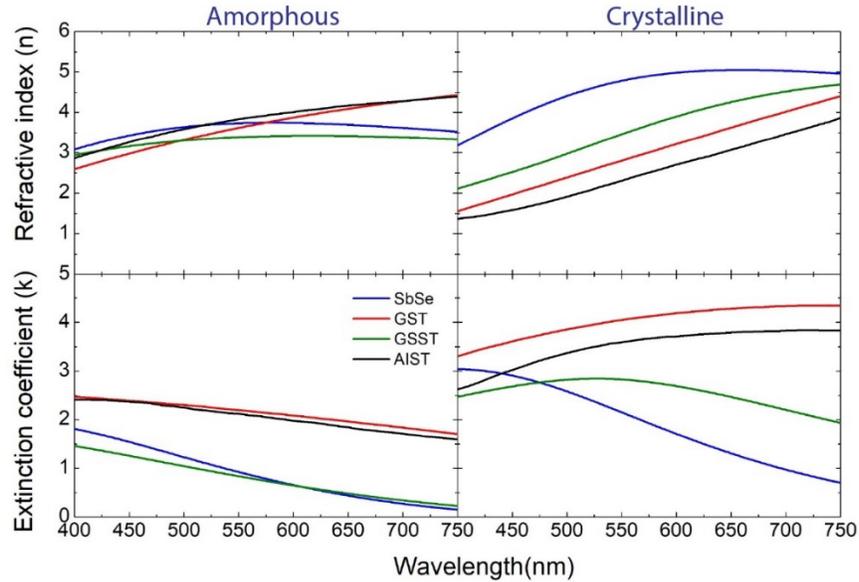

**Fig. S1. Refractive index of $Sb_2Se_3$ (SbSe), $Ge_2Sb_2Te_5$ (GST), $Ge_2Sb_2Se_4Te$ (GSST), $Ag_3In_4Sb_{76}Te_{17}$ (AIST).**

## S2. Effect of annealing process on the optical response of Ag/ITO/Ag

Fig. S2 shows the experimental results for the reflectance spectra before and after annealing. A strong effect in the optical response for wavelengths over 600 nm is attributed to dewetting in the thin-film silver layers. [2]

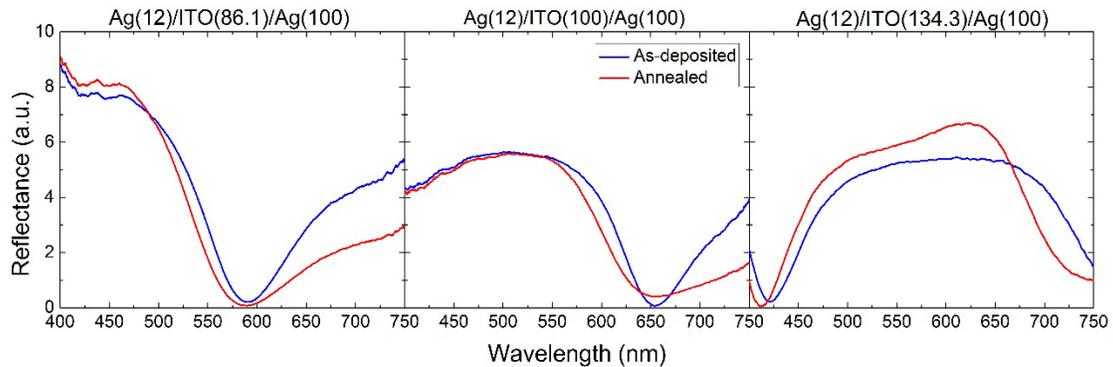

**Fig. S2. Effect of 200C for 10 min annealing process on the reflectance of different Ag/ITO/Ag structures.** The thicknesses on top of each figure are given in nanometers.

## S3. RGB color simulation of PCM-FROC devices

In **Fig. S3**, we demonstrate the color contrast for the sweeping thickness of both the $Sb_2Se_3$ and ITO layers. To be consistent with our device images in Fig 3 (main text). We include a 1 mm $SiO_2$ substrate in our simulations to consider the optical effect of the microscope glass slides onto which the sample is placed.

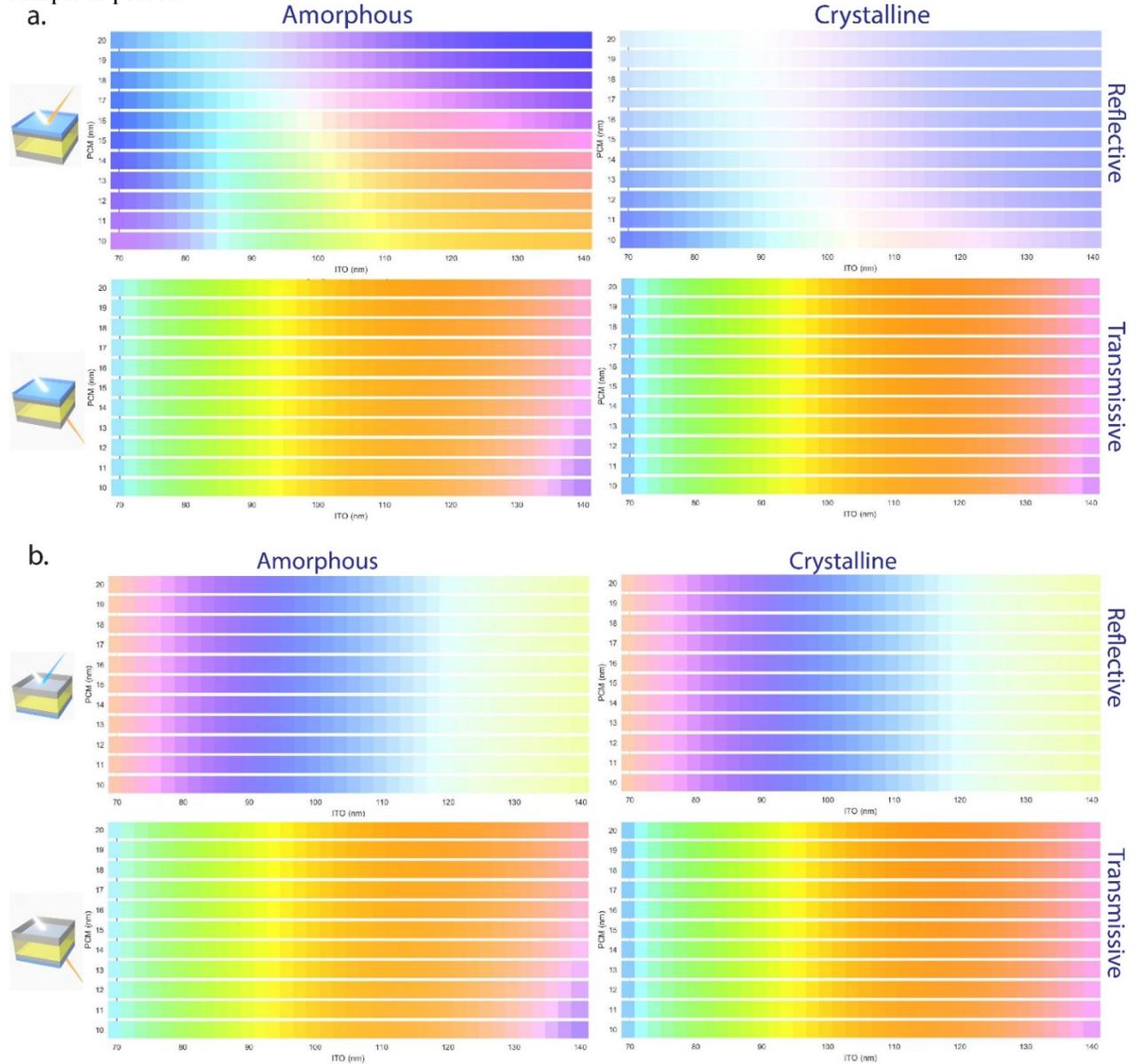

**Fig. S3. RGB color simulation of PCM-FROC devices**. **a.** Forward reflection and transmission color swatches for different thicknesses of $Sb_2Se_3$ from 10 to 20nm and ITO from 70 to 140nm. **b.** Backward reflection and transmission color swatches for $Sb_2Se_3$ thicknesses from 10 to 20nm and ITO from 70 to 140nm.

## S4. FDTD simulations of electric field in the PCM-FROC

In **Fig. S4**, we show the Lumerical FDTD simulations of the electric field distribution in the cross section of our PCM-FROCs with structure 17.3nm $Sb_2Se_3$/ 30nm Ag/ $t$ ITO/ 25nm Ag (see Fig. 3 in main text). In Fig. S3a, we show the simulation results considering light incidence from top, i.e., from the PCM side. we observe a strong resonance in the ITO layer at the same wavelength in which no electric field is localized in the $Sb_2Se_3$ layer (i.e., no absorption is observed, as described in [3]; this wavelength corresponds to the peak in reflection and transmission. Fig. S3b, on the other hand, show that upon backward illumination (i.e., light is incident on the bottom Ag layer), most wavelengths interact with the Ag-ITO-Ag Fabry-Perot cavity with a weak interaction with the $Sb_2Se_3$ layer. This effect is responsible for the narrow-band filter response upon backside illumination in Fig. 3 (main text).

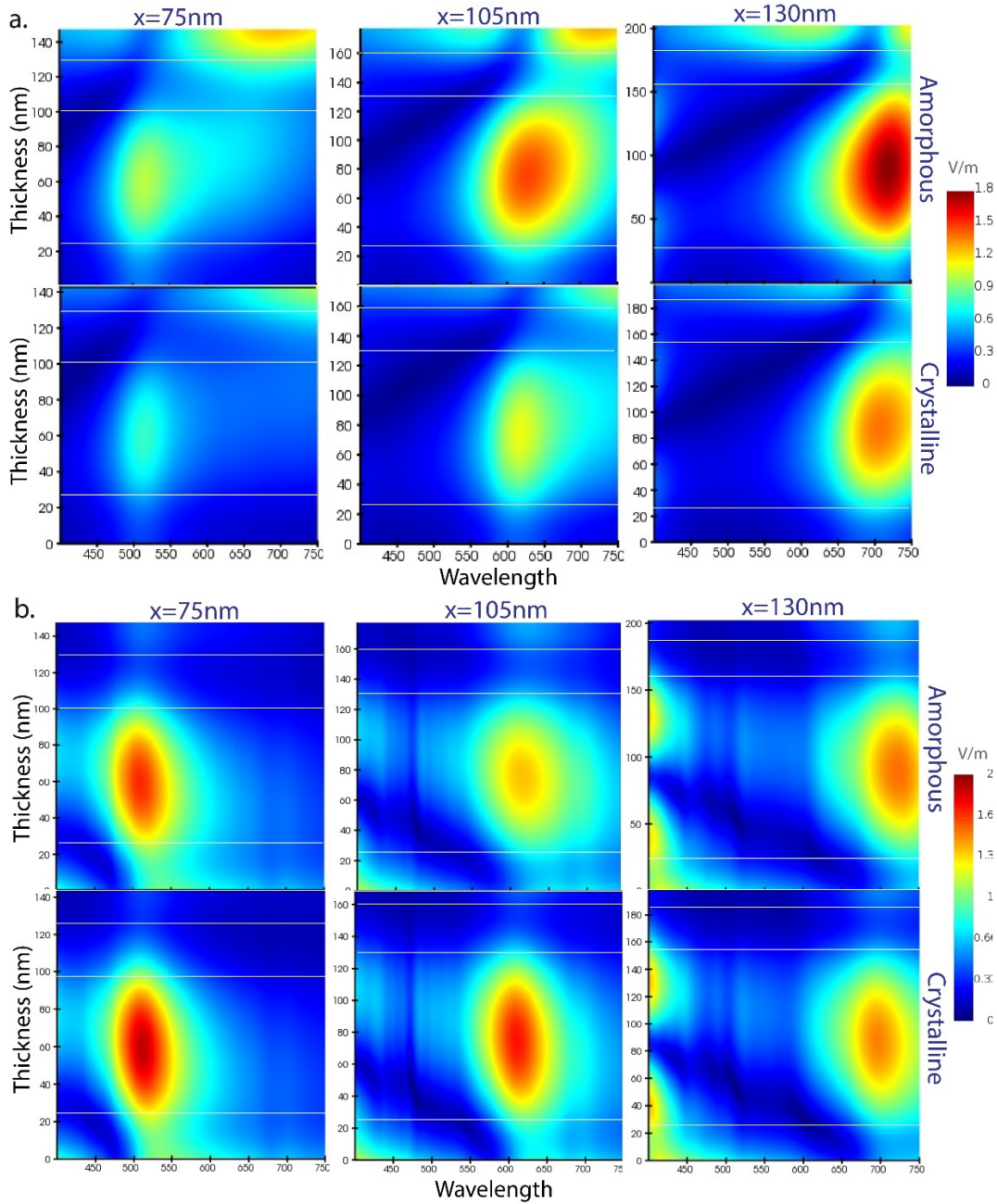

**Fig. S4. FDTD simulations of the electric field in the PCM-FROC cross section.** The horizontal lines show the boundaries between the $Sb_2Se_3$/Ag/ITO/Ag (from top to bottom.) **a.** FDTD simulations of Electric field with illumination on the $Sb_2Se_3$ side (top side). **b**. Same as a but with illumiation from the bottom Ag layer (bottom side).

## S5. Histogram of thickness distribution

Fig. S4 shows the thickness distribution for each of the four layers in a transmissive PCM-FROC with two objective optimizations in Fig. 5a. In the case of Sb$_2$Se$_3$, the solutions show a 6-7nm thin film solution for the second-order resonance centered around 400nm, while the thicker one, ~25nm, favor the red wavelengths with maximum amplitude modulation. The optimum solutions for the two Ag layers were considered in the fabrication of the samples.

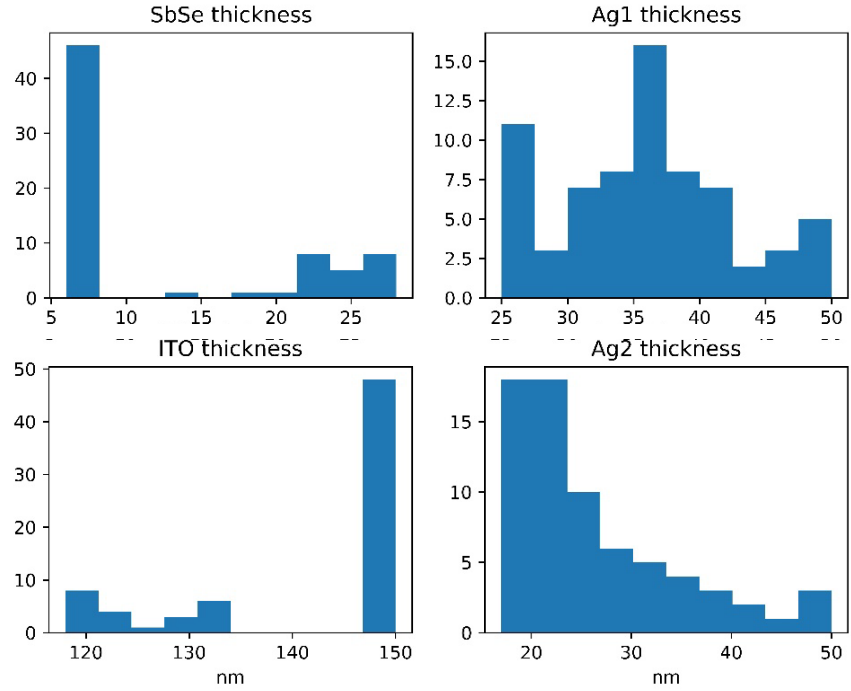

**Fig. S5. The histogram of thickness distribution within the solution sets in different layers.**

## S6. Four objectives optimization: FWHM and peak transmittance in Amorphous and Crystalline states

Multi-objective optimization for peaks' amplitude and FWHM for both the amorphous and the crystalline states. Most peaks in the solution sets with the optimized structures concentrate mainly around 400 nm and 750 nm. The ones located at 750 nm show the second resonances at 400 nm as the minor peaks. An important conclusion from these four objectives is that the results converged to similar structures as those found with two objectives optimization, i.e., only for the amorphous state (see Fig .5a). For this reason, we focused mostly on two objective simulations in the amorphous state in any other optimization task.

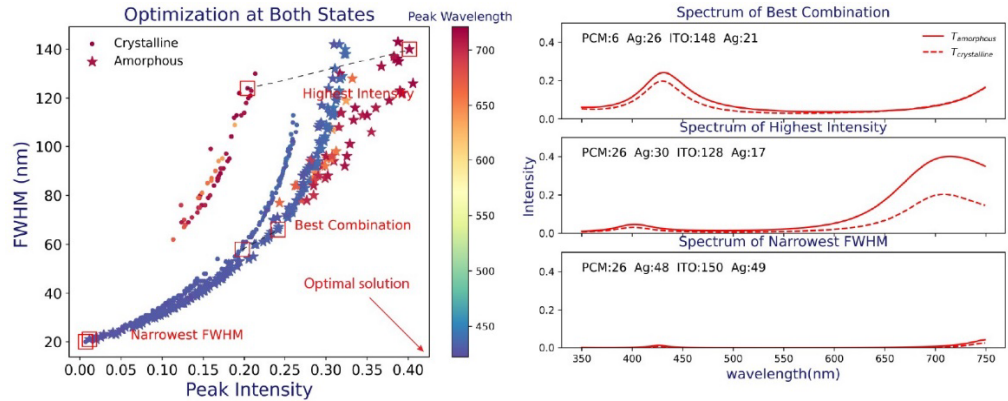

**Fig. S6.** Multi-objective optimization for peaks' amplitude and FWHM for the amorphous and the crystalline states. There are totally four objectives in this problem.

## S7. Three objectives optimization: FWHM, peak transmittance and peak shift

Multi-objective optimization in terms of peaks' amplitude and FWHM in the amorphous state, and peak shift between two states, for a total of three objectives in this problem. We observed that the maximum peak shift between amorphous and crystalline state is 20 nm when the thickness of PCM is thick (28 nm) and the main peaks of the spectra are located in short wavelengths. The best combination for this optimization task, shown in Fig. S6, refers to the PCM-FROC that allows for achieving the highest peak intensity with the smallest FWHM, i.e., towards the purest color. However , the best solution also leads to a case in which the peak shift upon crystallization is very low, displaying basically the same optical response regardless of the PCM state.

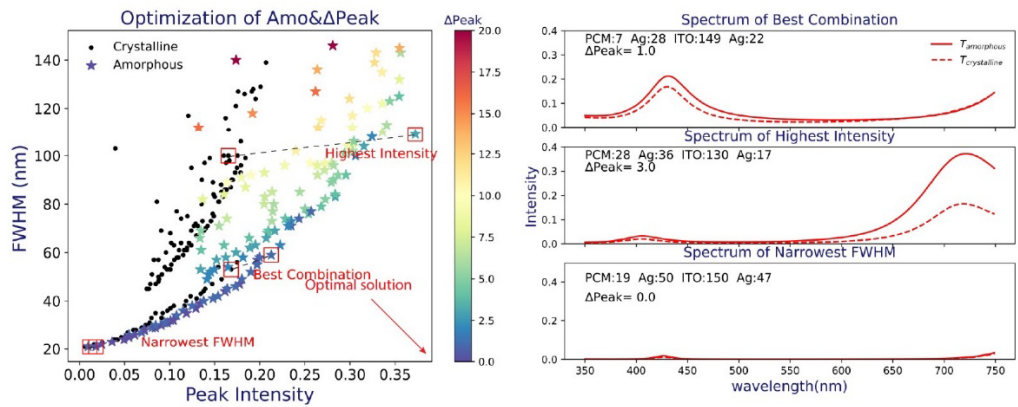

**Fig. S7.** Multi-objective optimization with three objectives: peaks' amplitude and FWHM in the amorphous state and peak shift between the two states.